\UseRawInputEncoding
\documentclass[12pt]{article}
\usepackage{amsfonts}
\usepackage{cite}
\usepackage{amsmath}
\usepackage{amssymb}
\usepackage{graphicx}

\providecommand{\U}[1]{\protect\rule{.1in}{.1in}}
\makeatletter
\@addtoreset{equation}{section}
\makeatother

\input{tcilatex}
\begin{document}

\begin{titlepage}
\vspace{.3cm}
\begin{center}
\renewcommand{\thefootnote}{\fnsymbol{footnote}}
{\Large{Flow Equations In Arbitrary Signature}}
\vskip1cm
\vskip1.3cm
W. A. Sabra 
\vskip1cm
\vskip.6cm {\small{\it
Physics Department, American University of Beirut\\ Lebanon  \\}}
\end{center}
\vfill\begin{center}
\textbf{Abstract}
\end{center}
\begin{quote}

We discuss general bosonic configurations of four-dimensional $N=2$ supergravity coupled to vector multiplets 
in $(t,s)$ space-time. The supergravity theories with Euclidean and neutral signature are 
described by the so-called para-special K\"{a}hler geometry. For extremal solutions, we derive in a unified fashion, using the equations of motion, the flow equations for all space-time signatures. 
Demanding that the solutions with neutral and Euclidean signatures admit unbroken supersymmetry, we derive the constraints, known as the stabilisation equations,  on the para-covariantly holomorphic sections expressed in terms of the adapted coordinates. The stabilisation equations expressed in terms of the para-complex sections imply generalised flow equations in terms of para-complex central charge. For Euclidean and neutral signature, it is demonstrated that solutions for either signs of gauge kinetic terms are mapped into each other via field redefinitions.
\end{quote}
\vfill\end{titlepage}

\section{Introduction}

The study of supersymmetric and BPS gravitational backgrounds of
supergravity theories in various space-time dimensions and signatures has
been an active area of research in recent years. Such backgrounds are of
importance for a better understanding of the non-perturbative structure of M
theory, the stringy duality symmetries as well as quantum geometry. The
first step in the systematic classification of solutions admitting
supersymmetry in Einstein-Maxwell theory was taken in \cite{Tod} building on
the earlier work of \cite{hullgib}. A result obtained is that supersymmetric
solutions with a time-like Killing vector are the well known IWP\ solutions 
\cite{IWP}. Later, a great deal of progress has been made in the study of
solutions of the more general Lorentzian $N=2$ four-dimensional supergravity
theories coupled with vector multiplets. Extensive details on $N=2$
supergravity with vector multiplets and their associated special K\"{a}hler
geometry can be found in \cite{speone}.

Supersymmetric solutions of the general $N=2$ supergravity theories were
first explored in \cite{origin}. A main result of \cite{origin} is that the
scalar fields of the supersymmetric solutions follow attractor equations
with fixed points on the horizon which are independent of their values at
infinity. The solutions at the near horizon are fully determined in terms of
algebraic conditions, the so-called stabilisation equations, 
\begin{equation}
i\left( \bar{Z}L^{I}-Z\bar{L}^{I}\right) =p^{I},\text{ \ \ \ \ \ }i\left( 
\bar{Z}M_{I}-\bar{Z}\bar{M}_{I}\right) =q_{I},  \label{cc}
\end{equation}
where $Z$ is the central charge and $p^{I}$\ and $q_{I}$ being the magnetic
and electric charges.

Static solutions with non-constant scalar fields were later constructed in 
\cite{germany}. The analogue of the stationary IWP solutions in the general
Lorentzian four-dimensional $N=2$ supergravity were first analysed in \cite%
{station}. The special K\"{a}hler geometry of the scalar fields proved to be
an essential ingredient in the analysis and the study of solutions admitting
supersymmetry. A rederivation of these solutions for either sign of the
gauge fields kinetic terms was performed in \cite{phantom} employing
spinorial geometry method as outlined in \cite{4spinor}. The solutions with
the standard sign of gauge kinetic terms are given by 
\begin{equation}
ds^{2}=-\frac{1}{|\beta |^{2}}\left( d\tau +\sigma \right) ^{2}+|\beta
|^{2}\left( dx^{2}+dy^{2}+dz^{2}\right) ,  \label{soul}
\end{equation}%
with the conditions 
\begin{eqnarray}
i\left( \bar{\beta}L^{I}-\beta \bar{L}^{I}\right) &=&\tilde{H}^{I},\text{ \
\ \ \ \ }i\left( \bar{\beta}M_{I}-\beta \bar{M}_{I}\right) =H_{I},
\label{stable} \\
d\sigma &=&\ast \left( H_{I}d\tilde{H}^{I}-\tilde{H}^{I}dH_{I}\right)
\end{eqnarray}%
where $\tilde{H}^{I}$ and $H_{I}$ are harmonic functions on the flat
coordinate space $\mathbb{R}^{3}$ and $\beta $ is a complex function$.$ Here 
$\ast $ is the Hodge star operator on flat $\mathbb{R}^{3}.$ Note that
solutions with non-canonical sign of gauge fields kinetic term have a
space-like Killing vector with the harmonic functions and the $\ast $ are on 
$\mathbb{R}^{1,2}$\cite{phantom}. Composite BPS configurations were
considered in \cite{den}. There, the correspondence between BPS states in
type II string theory compactified on Calabi-Yau manifolds and BPS\
solutions of the four-dimensional $N=2$ supergravity were investigated.

A thorough and detailed study of four-dimensional Euclidean supersymmetric
theories has recently been conducted in \cite{mohaupt1, mohaupt2, mohaupt3,
mohaupt4}. In particular, Euclidean $N=2$ supergravity theories were
obtained as dimensional reductions of five-dimensional $N=2$ supergravity
theories \cite{GST} on a time-like circle. The couplings of the Euclidean
theory were found to be described in terms of a para-special K\"{a}hler
geometry. More on the construction and the study of $N=2$ Euclidean
supergravity theories can be found in \cite{reduction, 5euclidean}.

A large class of Lorentzian four-dimensional $N=2$ supergravities are
obtainable from five-dimensional $N=2$ supergravity theory with $\left(
1,4\right) $ signature via a reduction on a space-like circle \cite{GST}. In
turn, the five-dimensional theories are reductions of $\left( 1,10\right) $
supergravity on a Calabi-Yau threefold $CY_{3}$ \cite{cad}\footnote{%
We use the notation $(t,s)$ for space-time signature where $t$ is the number
of time dimensions and $s$ is the number of spatial dimensions.}.
Four-dimensional supergravity theories with various space-time signatures
were obtained in \cite{cyr} by reducing the eleven-dimensional supergravity
theories \cite{Hull, cjs} on a $CY_{3}$ and a circle. A\ rigorous study of
the four-dimensional theories in various space-time signatures and the
classification of their four-dimensional $N=2$ supersymmetry algebras have
been performed in \cite{moh}.

In the present work we are mainly interested in the extension of the results
obtained for the solutions of the Lorentzian supergravity theories to all
four-dimensional supergravities with vector multiplets in various space-time
signatures. Our work is organized as follows. In section two we discuss
general solutions for all signatures following on the analysis of \cite{FWK}
. We derive flow equations for extremal solutions through the analysis of
the equations of motion. In section three, using spinorial geometry, we
analyse solutions for a specific choice of spinor orbit and construct the
corresponding IWP-like Euclidean and neutral signature solutions. We derive
the stabilisation equations in terms of the so-called adapted coordinates
and also in terms para-complex variables and derive generalised flow
equations. We summarize and conclude in section four.

\section{General Solutions and Flow equations}

Ignoring the hypermultiplets, the $N=2$ supergravity theories have the
bosonic part of their Lagrangian given by 
\begin{equation}
\mathbf{e}^{-1}\mathcal{L}=R-2g_{A\bar{B}}\partial _{\mu }z^{A}\partial
^{\mu }\bar{z}^{B}-\frac{\kappa ^{2}}{2}F^{I}\cdot \left( \func{Im}\mathcal{%
N }_{IJ}F^{J}+\func{Re}\mathcal{N}_{IJ}\tilde{F}^{J}\right) \;.  \label{4ac}
\end{equation}
For theories with Lorentzian signature, the fields $z^{A}$ are $n$ complex
scalar fields, $F^{I}$ and $\tilde{F}^{I}$ $(I=0,...,n)$ are two-forms
representing the gauge field-strengths and their duals and $\kappa ^{2}=\pm
1.$ The Lorentzian theories are described in terms of special K\"{a}hler
geometry which can be formulated in terms of the covariantly holomorphic
sections

\begin{equation}
V=\left( 
\begin{array}{c}
L^{I} \\ 
M_{I}%
\end{array}
\right) ,\text{ \ \ \ }I=0,...,n,\text{ \ \ }
\end{equation}%
satisfying 
\begin{eqnarray}
i\langle V,\bar{V}\rangle &=&i\left( \bar{L}^{I}M_{I}-L^{I}\bar{M}
_{I}\right) =1,  \label{sym} \\
\text{\ }D_{\bar{A}}V &=&\left( \partial _{\bar{A}}-\frac{1}{2}\partial _{ 
\bar{A}}K\right) V=0.
\end{eqnarray}
where $K$ is the K\"{a}hler potential. \ Defining%
\begin{equation}
U_{A}=D_{A}V=\left( \partial _{A}+\frac{1}{2}\partial _{A}K\right) V,
\end{equation}
we have the useful relations 
\begin{equation}
M_{I}=\mathcal{N}_{IJ}L^{J},\ \ \ \ D_{A}M_{I}=\mathcal{\bar{N}}
_{IJ}D_{A}L^{J}
\end{equation}
and 
\begin{eqnarray}
\langle D_{A}V,\bar{V}\rangle &=&\langle D_{A}V,V\rangle =\langle
D_{A}V,D_{B}V\rangle =0,  \notag \\
g_{A\bar{B}} &=&-i\langle D_{A}V,D_{\bar{B}}\bar{V}\rangle .  \label{ult}
\end{eqnarray}
The theories with Euclidean and neutral signature are formulated in terms of
para-special K\"{a}hler geometry. Roughly speaking, the equations of para
special K\"{a}hler geometry can be obtained from those of special K\"{a}hler
geometry by replacing $i$ with $\iota ,$ with $\bar{\iota}=-\iota $ and $\
\iota ^{2}=1$.

Following the analysis of \cite{FWK} for static solutions, we start by
considering the following metrics 
\begin{equation}
ds^{2}=\epsilon _{1}e^{2U}d\tau ^{2}+e^{-2U}\left( \epsilon _{0}{l}^{4}d\rho
^{2}+l^{2}ds^{2}(\mathcal{M}_{2})\right) ,  \label{ans}
\end{equation}
where $U$ and ${l}$ are functions of $\rho $ only and $\mathcal{M}_{2}$ is a
two-dimensional Einstein manifold with Ricci tensor 
\begin{equation}
\mathrm{R}_{ab}=kh_{ab}.
\end{equation}
The two-dimensional space $\mathcal{M}_{2}$ can be represented by the metric 
\begin{equation}
ds^{2}(\mathcal{M}_{2})=h_{ab}dx^{a}dx^{b}=\epsilon _{2}d\theta
^{2}+\epsilon _{3}f^{2}d\phi ^{2}.  \label{ec}
\end{equation}%
Here $\epsilon _{i}=\pm 1,$ $i=0,1,2,3.$ Euclidean and neutral signature
solutions have $\epsilon _{0}\epsilon _{1}\epsilon _{2}\epsilon
_{3}=\epsilon =1,$ while Lorentzian solutions have $\epsilon _{0}\epsilon
_{1}\epsilon _{2}\epsilon _{3}=\epsilon =-1.$ Moreover, the function $f$ can
take the values $\sin \theta $, $\sinh \theta $ or $1$. The manifold $%
\mathcal{M}_{2}$ can be of spherical $\left( k=1\right) $, hyperbolic $%
\left( k=-1\right) $ and flat topology $\left( k=0\right) $ depending on the
choice of values of $\epsilon _{2},\epsilon _{3}$ and $f$.

The Einstein equations of motion derived from the action (\ref{4ac}) are
given by 
\begin{equation}
R_{\mu \nu }=2g_{A\bar{B}}\partial _{\mu }z^{A}\partial _{\nu }\bar{z}
^{B}+2\kappa ^{2}\func{Im}\mathcal{N}_{IJ}\left( F{}_{\mu \lambda
}^{I}F^{J}{}_{\nu }{}^{\lambda }-{\frac{1}{4}}g_{\mu \nu }F^{I}{}_{\alpha
\beta }F^{J}{}^{\alpha \beta }\right) .  \label{ee}
\end{equation}%
The non-vanishing components of the Ricci-tensor of the metric (\ref{ans})
are 
\begin{align}
R_{\rho \rho }& =2\left( \partial _{\rho }\log l\right) ^{2}-2\partial
_{\rho }^{2}\log l+\partial _{\rho }^{2}U-2\left( \partial _{\rho }U\right)
^{2},  \notag \\
R_{\tau \tau }& =-\frac{\epsilon _{0}\epsilon _{1}}{l^{4}}e^{4U}\partial
_{\rho }^{2}U,  \notag \\
R_{ab}& =\left( k-\frac{\epsilon _{0}}{l^{2}}\left( \partial _{\rho
}^{2}\log l-\partial _{\rho }^{2}U\right) \right) h_{ab}\ .
\end{align}%
Thus one obtains from (\ref{ee}) 
\begin{align}
\left( \partial _{\rho }U\right) ^{2}+g_{A\bar{B}}\partial _{\rho
}z^{A}\partial _{\rho }\bar{z}^{B}-e^{2U}V& =\left( \partial _{\rho }\log {l}
\right) ^{2}-\partial _{\rho }^{2}\log {l,}  \notag \\
\partial _{\rho }^{2}U& =e^{2U}V,  \notag \\
\partial _{\rho }^{2}\log l& =k\epsilon _{0}l^{2},  \label{an}
\end{align}
where $V$ is given by 
\begin{equation}
V=-\frac{\epsilon _{1}\kappa ^{2}}{2}\func{Im}\mathcal{N}_{IJ}\left(
e^{-4U}F^{I}{}_{\tau \rho }F^{J}{}_{\tau \rho }{}-\frac{\epsilon }{f^{2}}
F^{I}{}_{\theta \phi }F_{\theta \phi }^{J}\right) .  \label{vee}
\end{equation}
The equations (\ref{an}) simplify further if we consider solutions with 
\begin{equation}
\left( \partial _{\rho }\log l\right) ^{2}-\partial _{\rho }^{2}\log l=c^{2},
\label{non}
\end{equation}
where the constant $c$ can be regarded as a non-extremality parameter. Then
( \ref{non}) together with (\ref{an}) imply the three possibilities $l=\frac{%
c }{\sinh c\rho },$ $l=\frac{c}{\cosh c\rho }$ and $l=e^{c\rho }$ for ${\
\epsilon _{0}}k=1,$ ${\epsilon _{0}}k=-1$ and $k=0,$ respectively. The
Einstein equations of motion, for various topologies of $\mathcal{M}_{2}$,
reduce to 
\begin{align}
\left( \partial _{\rho }U\right) ^{2}+g_{A\bar{B}}\partial _{\rho
}z^{A}\partial _{\rho }\bar{z}^{B}-{e}^{2U}{V}& ={c}^{2}{,}  \notag \\
\partial _{\rho }^{2}U& ={e^{2U}V.}  \label{eom}
\end{align}
After solving for the gauge fields, the potential $V$ takes the form 
\begin{equation}
V=\frac{\epsilon _{1}\kappa ^{2}}{2}\left[ \func{Im}\mathcal{N}^{MI}\left( 
\func{Re}\mathcal{N}_{MN}q_{I}p^{N}+\func{Re}\mathcal{N}_{IL}p^{L}\left(
q_{M}-\func{Re}\mathcal{N}_{MN}p^{N}\right) -q_{I}q_{M}\right) +\epsilon 
\func{Im}\mathcal{N}_{IJ}p^{I}p^{J}\right]  \label{pt}
\end{equation}
where $p$ and $q$ represent the magnetic and electric charges. The potential 
$V$ can be expressed as 
\begin{equation}
V=-\epsilon \epsilon _{1}\kappa ^{2}\left( \left\vert Z\right\vert ^{2}+g^{A 
\bar{B}}D_{A}ZD_{\bar{B}}\bar{Z}\right)
\end{equation}
with $Z$ being the central charge given by 
\begin{equation}
Z=L^{I}q_{I}-M_{I}p^{I},
\end{equation}
where $L^{I}$ and $M_{I}$ are para-complex for theories with neutral and
Euclidean signature and complex for Lorentzian theories. For extremal
spherically symmetric Lorentzian theories with ${\kappa ^{2}=-1,}$ $\epsilon
=\epsilon _{1}=-1,$ one reproduces the results of \cite{FWK} where two
gradient flow equations were obtained 
\begin{eqnarray}
\partial _{\rho }U &=&\pm e^{U}\left\vert Z\right\vert ,  \label{f} \\
\partial _{\rho }z^{A} &=&\pm 2e^{U}g^{A\bar{B}}\partial _{\bar{B}
}\left\vert Z\right\vert .  \label{fl}
\end{eqnarray}
These first order differential equations were also obtained in \cite{origin}
using supersymmetry. It can be shown that the flow equations (\ref{f}) and ( %
\ref{fl}) imply all the equations of motion. The plus sign possibility on
the right hand side of (\ref{f}) and (\ref{fl}) leads to physically
unacceptable solutions \cite{den}. The same analysis holds for extremal
Lorentzian solutions with non-canonical sign of gauge kinetic terms where we
have ${\kappa ^{2}=1,}$ $\epsilon =\epsilon _{0}=-1.$ For extremal Euclidean
and neutral solutions, taking ${\epsilon _{i}=-\kappa }^{2}=\epsilon =1,$
one also gets the relation 
\begin{equation}
V=\left\vert Z\right\vert ^{2}+g^{A\bar{B}}D_{A}ZD_{\bar{B}}\bar{Z}
\end{equation}%
and as such we obtain the same flow equations expressed in terms of the
para-complex central charge.

For $c^{2}\neq 0,$ one obtains non-extremal solutions. A general analysis
for the study of non-supersymmetric solutions in arbitrary dimensions and
metric signatures was recently given in \cite{jsnon}. We must note that the
non-extremal metric ansatz (\ref{ans}) is related by a coordinate
transformation to that used in \cite{jsnon} and \cite{MO5}.

\section{Supersymmetric Euclidean and Neutral Solutions}

In this section, we shall consider Euclidean as well as neutral signature
solutions with either sign of the gauge kinetic term in a unified setting. A
systematic analysis of Euclidean and neutral solutions for Einstein-Maxwell
theory were recently considered in \cite{pe}. The Killing spinor equations
for these theories are given by \cite{cyr}: 
\begin{align}
\left( \nabla _{\mu }-{\frac{1}{2}}A_{\mu }\gamma _{5}+\frac{\kappa }{4}
\gamma \cdot F^{I}\left( \mathrm{Im}L^{J}+\gamma _{5}\mathrm{Re}L^{J}\right)
(\mathrm{Im}\mathcal{N})_{IJ}\gamma _{\mu }\right) \varepsilon & =0,
\label{ks1} \\
\frac{\kappa }{2}(\mathrm{Im}\mathcal{N})_{IJ}\gamma \cdot F^{J}\left[ 
\mathrm{Im}(D_{\bar{B}}{\bar{L}}^{I}g^{A\bar{B}})+\gamma _{5}\mathrm{Re}(D_{ 
\bar{B}}{\bar{L}}^{I}g^{A\bar{B}})\right] \varepsilon +\gamma ^{\mu
}\partial _{\mu }\left( \mathrm{Re}z^{A}-\gamma _{5}\mathrm{Im}z^{A}\right)
\varepsilon \,& =0,\   \label{ks2}
\end{align}
with $\kappa =i$ or $\kappa =-1.$ Here $A$ is the $U(1)$ K\"{a}hler
connection given by 
\begin{equation*}
A=-\frac{\iota }{2}(\partial _{A}Kdz^{A}-\partial _{\bar{A}}Kd\bar{z}^{A}).
\end{equation*}
We start our analysis by considering metrics of the form 
\begin{equation}
ds^{2}=2\left( \mathbf{e}^{1}\mathbf{e}^{\bar{1}}+\eta ^{2}\mathbf{e}^{2} 
\mathbf{e}^{\bar{2}}\right) ,  \label{met}
\end{equation}%
where $\eta ^{2}=-1$ for solutions with $(2,2)$ space-time signature and $%
\eta ^{2}=1$ for Euclidean solutions. Using spinorial geometry, we shall
analyze equations (\ref{ks1}) and (\ref{ks2}) for the spinor orbit $%
\varepsilon =\lambda 1+\sigma e_{1}$, with real functions $\lambda $ and $%
\sigma $. The action of the Dirac matrices on the Dirac spinors is given by 
\begin{equation}
\gamma _{2}=\sqrt{2}\eta i_{e^{2}},\ \ \ \gamma _{\bar{2}}=\sqrt{2}\eta
e^{2}\wedge ,\ \ \ \gamma _{1}=\sqrt{2}i_{e^{1}},\ \ \gamma _{\bar{1}}=\sqrt{
2}e^{1}\wedge  \label{acc}
\end{equation}
where $\{1,e_{1},e_{2},e_{12}=e_{1}\wedge e_{2}\}$ is the basis space of
forms on $\mathbb{R}^{2}.$ Plugging $\varepsilon =\lambda 1+\sigma e_{1}$ in
(\ref{ks1}) and using (\ref{acc}), we obtain the following geometric
conditions on the spin connection 
\begin{eqnarray}
\omega _{1\bar{1}} &=&\left( \partial _{2}\log \frac{\lambda }{\sigma }
-A_{2}\right) \mathbf{e}^{2}-\partial _{1}\log \lambda \sigma \mathbf{e}
^{1}-\left( \partial _{\bar{2}}\log \frac{\lambda }{\sigma }-A_{\bar{2}
}\right) \mathbf{e}^{\bar{2}}+\partial _{\bar{1}}\log \lambda \sigma \mathbf{%
\ e}^{\bar{1}},  \notag \\
\omega _{2\bar{2}} &=&\eta ^{2}\left( \left( \partial _{\bar{1}}\log \frac{
\lambda }{\sigma }-A_{\bar{1}}\right) \mathbf{e}^{\bar{1}}-\left( \partial
_{1}\log \frac{\lambda }{\sigma }-A_{1}\right) \mathbf{e}^{1}+\partial
_{2}\log \lambda \sigma \mathbf{e}^{2}-\partial _{\bar{2}}\log \lambda
\sigma \mathbf{e}^{\bar{2}}\right) ,  \notag \\
\omega _{12} &=&\kappa ^{2}\left( \partial _{2}\log \lambda
^{2}-A_{2}\right) \mathbf{e}^{1}+\eta ^{2}\left( \partial _{1}\log \lambda
^{2}-A_{1}\right) \mathbf{e}^{\bar{2}},\text{ \ \ \ }  \notag \\
\text{\ }\omega _{\bar{1}2} &=&\kappa ^{2}\left( \partial _{2}\log \sigma
^{2}+A_{2}\right) \mathbf{e}^{\bar{1}}+\eta ^{2}\left( \partial _{\bar{1}
}\log \sigma ^{2}+A_{\bar{1}}\right) \mathbf{e}^{\bar{2}},
\end{eqnarray}
together with 
\begin{eqnarray}
\mathrm{Im}\mathcal{N}_{IJ}L_{-}^{I}\left( F_{1\bar{1}}^{J}-\eta ^{2}F_{2 
\bar{2}}^{J}\right) &=&-\frac{\kappa \sqrt{2}}{\sigma \lambda }\left(
\partial _{1}+A_{1}\right) \sigma ^{2},  \notag \\
\mathrm{Im}\mathcal{N}_{IJ}L_{+}^{I}\left( F_{1\bar{1}}^{J}+\eta ^{2}F_{2 
\bar{2}}^{J}\right) &=&\frac{\sqrt{2}\bar{\kappa}}{\sigma \lambda }\left(
\partial _{1}-A_{1}\right) \lambda ^{2},  \notag \\
\mathrm{Im}\mathcal{N}_{IJ}L_{-}^{I}F_{\bar{1}2}^{J} &=&\frac{\kappa }{\sqrt{
2}\lambda \sigma }\left( \partial _{2}+A_{2}\right) \sigma ^{2},  \notag \\
\mathrm{Im}\mathcal{N}_{IJ}L_{+}^{I}F_{12}^{J} &=&-\frac{\kappa }{\sqrt{2}
\lambda \sigma }\left( \partial _{2}-A_{2}\right) \lambda ^{2}.  \label{feq}
\end{eqnarray}%
where we have expressed our relations in terms of the so-called adapted
coordinates (real light-cone coordinates) \cite{mohaupt4} defined as $X_{\pm
}=\func{Re}X\pm \func{Im}X.$ The analysis of (\ref{ks2}) gives%
\begin{eqnarray}
\partial _{1}z_{-}^{A} &=&\frac{\kappa \sigma }{\sqrt{2}\lambda }\mathrm{Im} 
\mathcal{N}_{IJ}(D_{\bar{B}}{\bar{L}}^{I}g^{A\bar{B}})_{-}\left( F_{1\bar{1}
}^{J}-\eta ^{2}F_{2\bar{2}}^{J}\right) ,  \notag \\
\partial _{\bar{1}}z_{+}^{A} &=&\frac{\kappa \lambda }{\sqrt{2}\sigma } 
\mathrm{Im}\mathcal{N}_{IJ}\text{\ }(D_{\bar{B}}{\bar{L}}^{I}g^{A\bar{B}
})_{+}\left( F_{1\bar{1}}^{J}+\eta ^{2}F_{2\bar{2}}^{J}\right) ,  \notag \\
\partial _{2}z_{+}^{A} &=&\frac{\sqrt{2}\kappa \lambda }{\sigma }\mathrm{Im} 
\mathcal{N}_{IJ}\text{\ }(D_{\bar{B}}{\bar{L}}^{I}g^{A\bar{B}
})_{+}F_{12}^{J},  \notag \\
\partial _{2}z_{-}^{A} &=&-\frac{\sqrt{2}\kappa \sigma }{\lambda }\mathrm{Im}
\mathcal{N}_{IJ}(D_{\bar{B}}{\bar{L}}^{I}g^{A\bar{B}})_{-}F_{\bar{1}2}^{J}.
\label{zeq}
\end{eqnarray}
One also obtains 
\begin{equation}
\left( \partial _{\bar{1}}+\kappa ^{2}\partial _{1}\right) \lambda =\left(
\partial _{\bar{1}}+\kappa ^{2}\partial _{1}\right) \sigma =\left( \partial
_{1}+\kappa ^{2}\partial _{\bar{1}}\right) z_{\pm }^{A}=\left( A_{\bar{1}
}+\kappa ^{2}A_{1}\right) =0.
\end{equation}
The torsion free condition implies that 
\begin{eqnarray}
d\left( \mathbf{e}^{1}-\kappa ^{2}\mathbf{e}^{\bar{1}}\right) &=&-d\log
\lambda \sigma \wedge \left( \mathbf{e}^{1}-\kappa ^{2}\mathbf{e}^{\bar{1}
}\right)  \notag \\
d\mathbf{e}^{2} &=&-d\log \lambda \sigma \wedge \mathbf{e}^{2}.  \label{co}
\end{eqnarray}
All the above conditions indicates that we can introduce the coordinates $%
\tau ,$ $x,$ $y$, and $z,$ and write 
\begin{equation}
\mathbf{e}^{1}=-\frac{\kappa }{\sqrt{2}}\left( \lambda \sigma (d\tau +\phi
)+ \frac{1}{\lambda \sigma }idx\right) ,\text{ \ \ \ }\mathbf{e}^{2}=\frac{1%
}{ \sqrt{2}\lambda \sigma }\left( dy+idz\right) \ .  \label{coordinates}
\end{equation}%
The metric solution obtained from (\ref{met}) is independent of the
coordinate $\tau $, and is given by 
\begin{equation}
ds^{2}=\left( \lambda \sigma \right) ^{2}(d\tau +\phi )^{2}+\frac{1}{\left(
\lambda \sigma \right) ^{2}}\left( dx^{2}+\eta ^{2}\left(
dy^{2}+dz^{2}\right) \right) \   \label{me}
\end{equation}%
with%
\begin{equation}
d\phi =\frac{2}{\left( \lambda \sigma \right) ^{2}}\ast _{3}\left( d\log 
\frac{\lambda }{\sigma }-A\right) .  \label{dphi}
\end{equation}%
Here $\ast _{3}$ is the Hodge dual defined with respect to the metric $%
\left( dx^{2}+\eta ^{2}\left( dy^{2}+dz^{2}\right) \right) .$ Our
orientation is such that $\epsilon _{1\bar{1}2\bar{2}}=\eta ^{2}.$ Using ( %
\ref{feq}), (\ref{zeq}), (\ref{coordinates}), (\ref{dphi}) and the relations
of the para-special K\"{a}hler geometry 
\begin{eqnarray}
g^{A\bar{B}}D_{A}L^{I}D_{\bar{B}}\bar{L}^{J} &=&\frac{1}{2}\left( \func{Im} 
\mathcal{N}\right) ^{IJ}-\bar{L}^{I}L^{J}, \\
dM_{I} &=&\mathcal{\bar{N}}_{IJ}dL^{J}+2A\func{Im}\mathcal{N}_{IJ}L^{J},
\end{eqnarray}
we obtain after some calculation 
\begin{eqnarray}
F^{I} &=&d\left[ \left( \sigma ^{2}L_{+}^{I}-\kappa ^{2}\lambda
^{2}L_{-}^{I}\right) (d\tau +\phi )\right] +\ast d\left( \frac{\kappa
^{2}L_{-}^{I}}{\sigma ^{2}}+\frac{L_{+}^{I}}{\lambda ^{2}}\right) , \\
G_{I} &=&d\left[ \left( \sigma ^{2}M_{I+}-\kappa ^{2}\lambda
^{2}M_{I-}\right) (d\tau +\phi )\right] +\ast d\left( \kappa ^{2}\frac{%
M_{I-} }{\sigma ^{2}}+\frac{M_{I+}}{\lambda ^{2}}\right) ,
\end{eqnarray}
where $G_{I}=\func{Re}\mathcal{N}_{IJ}F^{I}+\func{Im}\mathcal{N}_{IJ}\tilde{%
F }^{I}$. The Bianchi identities together with Maxwell equations 
\begin{equation}
dF^{I}=dG_{I}=0,
\end{equation}
give 
\begin{equation}
\nabla ^{2}\left( \frac{L_{+}^{I}}{\lambda ^{2}}+\frac{\kappa ^{2}L_{-}^{I}}{
\sigma ^{2}}\right) =0\text{, \ \ \ \ \ }\nabla ^{2}\left( \frac{M_{+I}}{
\lambda ^{2}}+\frac{\kappa ^{2}M_{-I}}{\sigma ^{2}}\right) =0,
\end{equation}
where $\nabla ^{2}=\partial _{x}^{2}+\eta ^{2}\left( \partial
_{y}^{2}+\partial _{z}^{2}\right) ,$ thus one obtains the stabilisation
equations 
\begin{equation}
\frac{L_{+}^{I}}{\lambda ^{2}}+\frac{\kappa ^{2}L_{-}^{I}}{\sigma ^{2}}= 
\tilde{H}^{I}\text{, \ \ \ \ \ }\frac{M_{+I}}{\lambda ^{2}}+\frac{\kappa
^{2}M_{-I}}{\sigma ^{2}}=H_{I}.  \label{stab}
\end{equation}%
Here $\tilde{H}^{I}$ and $H_{I}$ are harmonic functions in the
three-dimensional space with metric $dx^{2}+\eta ^{2}\left(
dy^{2}+dz^{2}\right) $. Furthermore, (\ref{stab}) imply the relations 
\begin{eqnarray}
A &=&-\frac{\kappa ^{2}}{2}\left( \lambda \sigma \right) ^{2}\left( H_{I}d 
\tilde{H}^{I}-\tilde{H}^{I}dH_{I}\right) +d\log \frac{\lambda }{\sigma }, \\
d\phi &=&\kappa ^{2}\ast _{3}\left( H_{I}d\tilde{H}^{I}-\tilde{H}
^{I}dH_{I}\right) \ .
\end{eqnarray}
In what follows, we recast the solutions in terms of para-complex functions.
For definiteness, we consider the case with $\kappa ^{2}=-1$ and introduce
the para-complex functions 
\begin{equation}
\mathbb{Y}^{I}=\bar{\beta}L^{I},\ \ \ \ \ \mathbb{F}_{I}=\bar{\beta}M_{I},
\end{equation}
where $L^{I},$ $M_{I}$ and $\beta $ are all para-complex function. The
metric (\ref{me}) then takes the form 
\begin{equation*}
ds^{2}=\frac{1}{|\beta |^{2}}(d\tau +\omega )^{2}+|\beta |^{2}\left(
dx^{2}+\eta ^{2}\left( dy^{2}+dz^{2}\right) \right) .
\end{equation*}
The function $\beta $ can be related to $\lambda $ and $\sigma $ as 
\begin{equation}
\beta =\frac{1}{2\lambda ^{2}\sigma ^{2}}\left[ \left( \lambda ^{2}+\sigma
^{2}\right) +\iota \left( \lambda ^{2}-\sigma ^{2}\right) \right] .
\label{beta}
\end{equation}
The symplectic constraint (\ref{sym}) implies 
\begin{equation}
e^{-2U}=|\beta |^{2}=\iota (\mathbb{\bar{Y}}^{I}\mathbb{F}_{I}-\mathbb{Y}%
^{I} \mathbb{\bar{F}}_{I})  \label{mss}
\end{equation}%
and the stabilisation equations (\ref{stab}) take the familiar form 
\begin{equation}
\iota \left( \mathbb{Y}^{I}-\mathbb{\bar{Y}}^{I}\right) =\tilde{H}^{I}\text{
, \ \ \ \ \ \ \ \ }\iota \left( \mathbb{F}_{I}-\mathbb{\bar{F}}_{I}\right)
=H_{I}.  \label{gs}
\end{equation}
which can be written in a more compact form 
\begin{equation}
2\func{Im}\left( \bar{\beta}V\right) =H\text{, \ \ \ \ \ \ \ \ }H=\left( 
\begin{array}{c}
\tilde{H}^{I} \\ 
H_{I}%
\end{array}
\right) .  \label{compact}
\end{equation}
Using the relations (\ref{ult}) (for para-complex variables) together with 
\begin{equation*}
dV=D_{A}Vdz^{A}+\iota AV,
\end{equation*}
we obtain from (\ref{compact}) the relations 
\begin{eqnarray}
Zd\rho &=&\left( d-\iota A\right) \beta ,  \label{qa} \\
\partial _{\rho }z^{A}\ \ \ &=&-e^{U}g^{A\bar{B}}e^{\iota \gamma }D_{\bar{B}
}\bar{Z}
\end{eqnarray}
where 
\begin{equation}
\beta =e^{-U}e^{\iota \gamma }=e^{-U}\left( \cosh \gamma +\iota \sinh \gamma
\right) .
\end{equation}
Note that the relation (\ref{qa}) implies 
\begin{eqnarray}
\partial _{\rho }U &=&-\func{Re}\left( \frac{Z}{\beta }\right) , \\
\left( d\gamma -A\right) &=&\func{Im}\left( \frac{Z}{\beta }\right) d\rho .
\end{eqnarray}
The flow equations discussed in the previous section are obtained by setting 
\begin{equation}
A=d\gamma .
\end{equation}

Inspecting the stabilisation equations as expressed in terms of the adapted
coordinates in (\ref{stab}) and following on the arguments presented in \cite%
{5euclidean}, it can be seen that the solution corresponding to $\kappa
^{2}=-1$ can be mapped into the solution with $\kappa ^{2}=1,$ via the
symmetry 
\begin{eqnarray}
M_{-I}^{\prime } &=&-L_{-}^{I},\text{ \ \ \ }M_{+I}^{\prime }=L_{+}^{I}, \\
L_{-}^{\prime I} &=&-M_{-I},\text{ \ \ \ }L_{+}^{\prime I}=M_{+I},
\end{eqnarray}
together with interchanging magnetic with electric charges. Here the prime
coordinates represent the fields in $\kappa ^{2}=1$ theory. Note that in
terms of para-complex sections, this transformation reads 
\begin{equation}
\left( 
\begin{array}{c}
L^{\prime I} \\ 
M_{I}^{\prime }%
\end{array}
\right) =\iota \left( 
\begin{array}{cc}
\mathrm{0} & \mathrm{1} \\ 
\mathrm{1} & \mathrm{0}%
\end{array}
\right) \left( 
\begin{array}{c}
L^{I} \\ 
M_{I}%
\end{array}
\right)
\end{equation}
which together with electric-magnetic duality imply 
\begin{equation}
Z^{\prime }=-\iota Z.
\end{equation}
Note also that 
\begin{equation}
\iota \left( \bar{L}^{\prime I}M_{I}^{\prime }-L^{\prime I}\bar{M}%
_{I}^{\prime }\right) =\iota \left( \bar{L}^{I}M_{I}-L^{I}\bar{M}_{I}\right)
=1.
\end{equation}%
It is also evident from (\ref{stab}) that solutions for $\kappa ^{2}=1$
(with primed variables) can be obtained form the solutions corresponding to $%
\kappa ^{2}=-1,$ via the field redefinitions 
\begin{equation}
L_{+}^{\prime I}=L_{+}^{I},\text{\ \ \ \ \ }L_{-}^{\prime I}=-L_{-}^{I}, 
\text{ \ \ \ \ }M_{+I}^{\prime }=M_{+I},\text{ \ \ }M_{-I}^{\prime }=-M_{-I}.
\end{equation}
This is the transformation discussed in \cite{moh}. In terms of the
para-complex variables this is given by 
\begin{equation}
\left( 
\begin{array}{c}
L^{\prime I} \\ 
M_{I}^{\prime }%
\end{array}
\right) =\iota \left( 
\begin{array}{cc}
\mathrm{1} & \mathrm{0} \\ 
\mathrm{0} & \mathrm{1}%
\end{array}
\right) \left( 
\begin{array}{c}
L^{I} \\ 
M_{I}%
\end{array}
\right)  \label{mo}
\end{equation}
and it implies that

\begin{equation}
Z^{\prime }=\iota Z.
\end{equation}
In this case we have 
\begin{equation}
\iota \left( \bar{L}^{\prime I}M_{I}^{\prime }-L^{\prime I}\bar{M}
_{I}^{\prime }\right) =-\iota \left( \bar{L}^{I}M-L^{I}\bar{M}_{I}\right)
=-1.
\end{equation}
The transformation (\ref{mo}) is local and connects the Lagrangians of
theories with either sign of gauge kinetic terms \cite{moh}. As explained in 
\cite{moh}, the transformation is induced by an isomorphism between the two $%
N=2$ superalgebras arising from the dimensional reduction of the
five-dimensional supersymmetry algebras.

As examples, we construct some explicit solutions corresponding to the
minimal models defined by 
\begin{equation}
M_{I}=Q_{IJ}L^{J}
\end{equation}
with $Q_{IJ}$ being the elements of a constant (para)-complex symmetric
matrix. The stabilisation equations (\ref{gs}) lead to the solutions 
\begin{equation*}
\mathbb{Y}^{K}=\bar{\beta}L^{K}=-\frac{1}{2}\func{Im}Q^{IK}\left( \bar{Q}
_{IJ}\tilde{H}^{J}-H_{I}\right) ,\ 
\end{equation*}%
We then obtain from (\ref{mss}) 
\begin{equation}
e^{-2U}=\frac{1}{2}H^{t}XH
\end{equation}
where 
\begin{equation}
X=\left( 
\begin{array}{cc}
\func{Re}Q\left( \func{Im}Q\right) ^{-1}\func{Re}Q-\func{Im}Q & -\func{Re}
Q\left( \func{Im}Q\right) ^{-1} \\ 
-\left( \func{Im}Q\right) ^{-1}\func{Re}Q & \left( \func{Im}Q\right) ^{-1}%
\end{array}
\right) .
\end{equation}
Next we can construct supersymmetric solutions for theories with cubic
prepotential which can be obtained from eleven dimensions via a
compactification on a $CY_{3}$ and a circle. These theories can be defined
by the relations%
\begin{equation}
M_{0}=-C_{IJK}\frac{L^{I}L^{J}L^{K}}{\left( L^{0}\right) ^{2}},\text{ \ \ \
\ }M_{I}=3C_{IJK}\frac{L^{J}L^{K}}{L^{0}}.
\end{equation}
The stabilization equations, for the specific choice $\tilde{H}^{0}=H_{I}=0,$
can be solved by

\begin{equation}
\mathbb{Y}^{I}=\frac{1}{2}\iota \tilde{H}^{I},\text{ \ \ \ \ \ \ \ }\mathbb{%
Y }^{0}=\frac{1}{2}\text{\ \ }\sqrt{-\frac{C_{IJK}\tilde{H}^{I}\tilde{H}^{J} 
\tilde{H}^{K}}{H_{0}}}
\end{equation}
and we obtain 
\begin{equation}
e^{-2U}=2\text{\ }\sqrt{-C_{IJK}H_{0}H^{I}H^{J}H^{K}}.
\end{equation}

\section{Final Remarks}

In this paper we have considered solutions to four-dimensional supergravity
theories in arbitrary space-time signature. The first order differential
flow equations, in terms of the (para)-complex central charge, for the
metric and scalar fields were derived via the analysis of the equations of
motion. The spinorial geometry methods were employed in the analysis of the
Killing spinor equations and a class of supersymmetric Euclidean and neutral
IWP-like solutions were derived. The solutions and in particular the
stabilisation equations characterizing the evolution of the scalar fields
were expressed in terms of adapted coordinates (real light-cone
coordinates). The stabilisation conditions expressed in terms of
para-complex variables were used to find the generalised flow equations
characterizing our supersymmetric solutions. For different signs of the
gauge kinetic terms, the solutions were related to each other via a field
redefinition in line with the discussions in \cite{5euclidean, moh}.

\ BPS states in type II theory compactified on $CY_{3}$ can be described in
terms of branes wrapped on various supersymmetric cycles. To match the two
spectrums, multi-centered composite solutions, corresponding to
multi-centered harmonic functions, must be taken into consideration \cite%
{den}. Some explicit solutions of such composite solutions were also given
in \cite{ec}. It would be of interest to perform similar analysis for
Euclidean and neutral solutions in relation to the various $CY_{3}$
compactifications of type II string theories obtained as circle reductions
of M-theory with space-time signatures $(1,10)$,($2,9)$ and $(5,6)$ and
their mirrors \cite{Hull}. It is also of interest to generalise the results
of \cite{pe} to include vector multiplets and thus have a complete
classification of supersymmetric solutions. Another important direction is
the construction of supergravity theories with higher derivative terms in
Euclidean and neutral signatures and the study of their BPS\ solutions. We
hope to report on some of these research directions in future publications.

\bigskip

\textbf{Acknowledgements}: The work is supported in part by the National
Science Foundation under grant number PHY-1620505.

\end{document}